\documentclass{evn2004}
\usepackage{txfonts}
\usepackage{natbib}
\usepackage{graphicx}
\begin{document}
   \title{Looking for prematurely `dying', young, compact radio sources}

   \author{M. Kunert-Bajraszewska \inst{1},
          A. Marecki \inst{1}
          \and
          R. E. Spencer \inst{2}
          }

   \institute{Toru\'n Centre for Astronomy, N. Copernicus University, Toru\'n,
             Poland
             \and
             Jodrell Bank Observatory, University of Manchester,
             Macclesfield, Cheshire, SK11 9DL, UK
             }
   \authorrunning{Kunert-Bajraszewska et al.}
\date{}

   \abstract
{
We present VLBA 1.6, 5, 8.4 and 15\,GHz observations of a new sample
of weak, young, compact candidates for radio faders selected from
the VLA FIRST survey. We claim that some, or even the majority of young
sources, may be short-lived phenomena due to a lack of stable fuelling
from the black hole and fade before evolving to large extended objects.
}

   \maketitle

\section{Introduction}

The activity period for radio-loud AGNs (RLAGN) can last up to $\sim
10^{8}$ years \citep{al87,liu92} and, as their lobes are huge reservoirs
of energy, even if the energy supply from the central engine to the
hotspots and the lobes eventually cuts off, the radio sources are still
observable for a substantial period of time. This so-called `coasting
phase' of the lobes of a RLAGN can last up to $10^{8}$\,yr
\citep{kom94,sl01} and preserves the information of past nuclear activity.
As the source gradually fades out, its spectrum becomes steeper and
steeper because of radiation and expansion losses. Objects possessing
these features are sometimes termed `faders'.

Many examples of diffuse radio emission have already been found in
clusters of galaxies \citep[see e.g.][]{dew91,har93,giov99,gov01,mur04}
and this number is still increasing. These objects have been identified as
radio relics taking the form of e.g. diffuse lobes without hotspots or
radio `haloes'. \citet{cor87} has also described the structure and
properties of a double radio source B2\,0924+30 as a possible relic radio
galaxy. The projected linear size of the whole system is 
375\,kpc\footnote{Throughout this paper the following cosmological
parameters have been adopted: $H_0$=72${\rm\,km\,s^{-1}\,Mpc^{-1}}$
\citep{hstkp01} and $q_0$=0.5.}
which
indicates that this source is a `dying' large symmetric object (LSO). A
question that naturally arises is whether activity periods can be shorter
than for those in LSOs.

\citet{r96} proposed an evolutionary scheme unifying three classes of
objects: Compact Symmetric Objects (CSOs), Compact Steep Spectrum sources
(CSSs) and Large Symmetric Objects (LSOs). One of the main arguments in
favour of this hypothesis is that CSOs and some CSS sources, namely
Medium-sized Symmetric Objects (MSOs), which are unbeamed CSSs, have
similar morphologies to LSOs. \citet{sn99,sn00} discussed many aspects of
the above scenario in detail. In particular they concluded that the radio
luminosities of CSO objects increase as they evolve, reach a maximum in
the CSS/MSO phase and then gradually decrease as these objects grow
further to become LSOs. \citet{mar46} claim that this evolutionary track
is not the only one possible. In fact, a whole family (a continuum?) of
such tracks might exist and the one shown by \citet{sn99,sn00} just {\em
appears} as the only one, simply because of selection effects. If the
energy supply cuts off earlier, the object leaves that `main sequence'
proposed by \citet{sn99,sn00} and it will never reach the LSO stage (at
least in that phase of activity). Thus, there should exist a class of
small-scale objects that resemble large-scale faders.

Strong support for such an idea comes from \citet{rb97}. They proposed a
model in which extragalactic radio sources are intermittent on timescales
of $\sim$$10^4$\,--\,$10^5$~years. When the power supply cuts off, the
radio source fades rapidly in radio luminosity. However, the shocked
matter continues to expand supersonically and keeps the basic source
structure intact. This model predicts that there should be a large number
of weaker MSOs than those known so far because of the power cut-off. The
findings of \citet{rb97} and \citet{mar46} clearly suggest that compact
faders may exist. The main goal of our observations has therefore been the
study of the evolution of RLAGNs and the location of {\em weak} CSSs in
the evolutionary scheme of radio sources.

\section{The observations}

Using the VLA FIRST catalogue we have selected a sample of 60 candidates
which could be weak Compact Steep Spectrum sources. The selection criteria
have been given by \citet{kun02}. All the sources were initially observed
with MERLIN at 5\,GHz. The results of these observations led to the
selection of several groups of objects for further study with VLBI
\citep{mar42}. One of those groups contains 6~sources (see
Table~\ref{table1}) that have been completely unresolved by MERLIN at
5\,GHz but still have very steep ($\alpha \leq -0.7$) spectra between 1.6
and 5\,GHz. We think that the nature of such sources can be explained
using the theoretical framework outlined in the previous section. If the
main evolutionary sequence proposed by \citet{sn99,sn00} is, as we claim
not the only one, potentially every steep spectrum source, could be a
candidate for a fader regardless of its size --- i.e. even the most
compact CSOs {\em could} be dying if they have plain steep spectra and
not Gigahertz-Peaked Spectra as CSOs normally do.

Therefore, we decided to observe the 6~unresolved sources mentioned above
with the VLBA at 1.6, 5, 8.4 and 15\,GHz in a snapshot mode. In a first
set of observations at 1.6\,GHz, the Effelsberg telescope was included in
the VLBA in order to improve the resolution at that relatively low
frequency.  The successful outcome from these observations led to a second
set of observations at the higher frequencies quoted above.

\begin{table}
\caption[]{Names and coordinates of target sources (J2000)}
\begin{tabular}{@{}c c c@{}}
\hline
\hline
       &    &    \\
~~Source & RA & DEC \\
~~Name   & h~m~s & $\degr$~$\arcmin$~$\arcsec$ \\
~~(1)& (2)& (3)~~~\\
\hline
\hline
~~0809+404 &08 12 53.124 &40 18 59.878~~\\
~~0949+287 &09 52 06.091 &28 28 32.406~~\\
~~1159+395 &12 01 49.965 &39 19 11.023~~\\
~~1315+396 &13 17 18.635 &39 25 28.141~~\\
~~1502+291 &15 04 26.696 &28 54 30.548~~\\
~~1616+366 &16 18 23.581 &36 32 01.811~~\\
\hline
\end{tabular}
\label{table1}
\end{table}

\section{Results and comments on 0809+404}

The whole data reduction process from the initial editing and calibration
to the production of final images was carried out using AIPS. Flux densities
of the principal components of the sources were measured using JMFIT and
their spectral indices from 1.6 to 4.9\,GHz, 4.9 to 8.4\,GHz and 8.4 to
15.4\,GHz were calculated. It is to be noted that only three sources were
detected at 15.4\,GHz.

\begin{figure*}[t]
\centering
\includegraphics[scale=0.3]{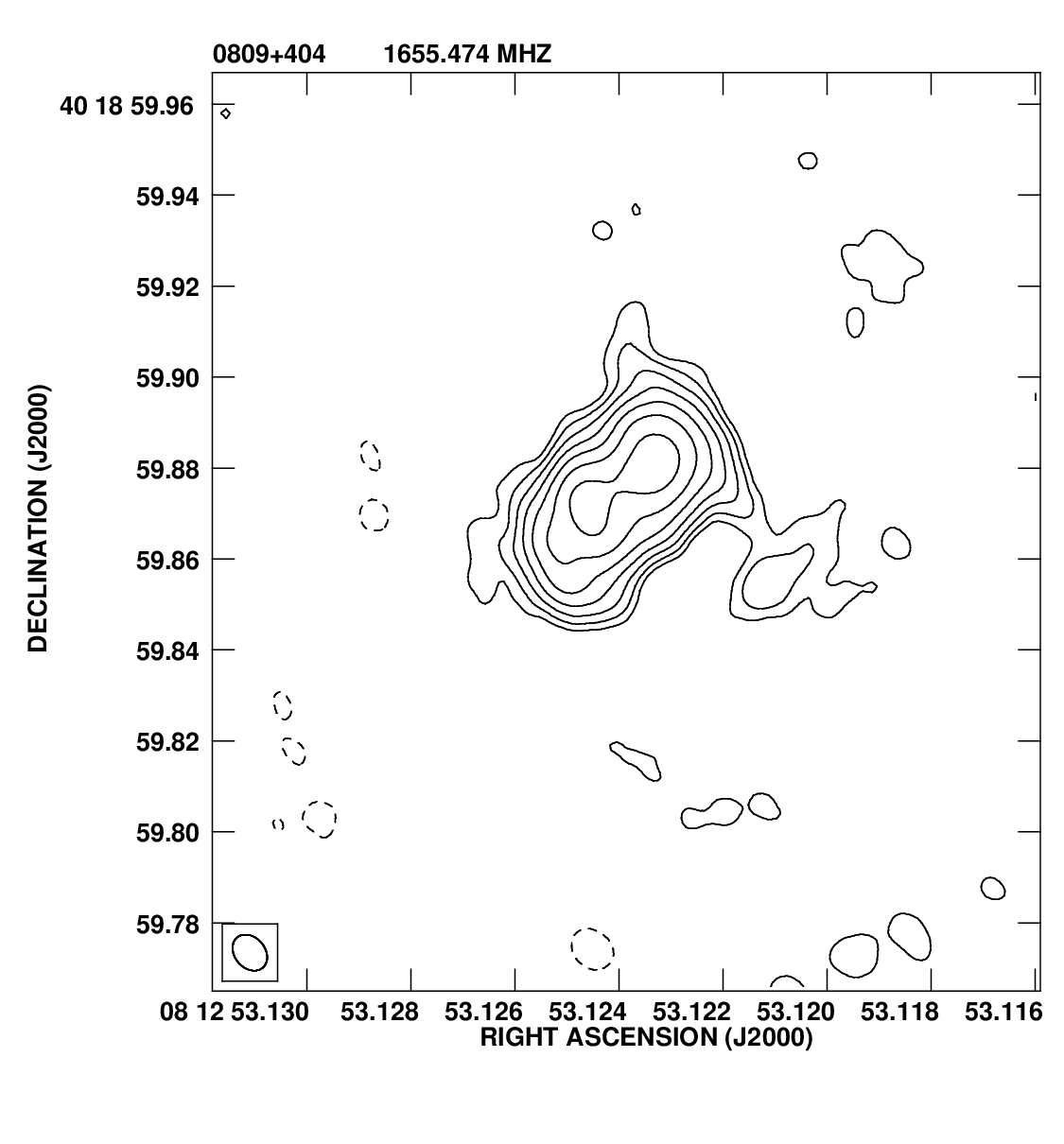}
\includegraphics[scale=0.3]{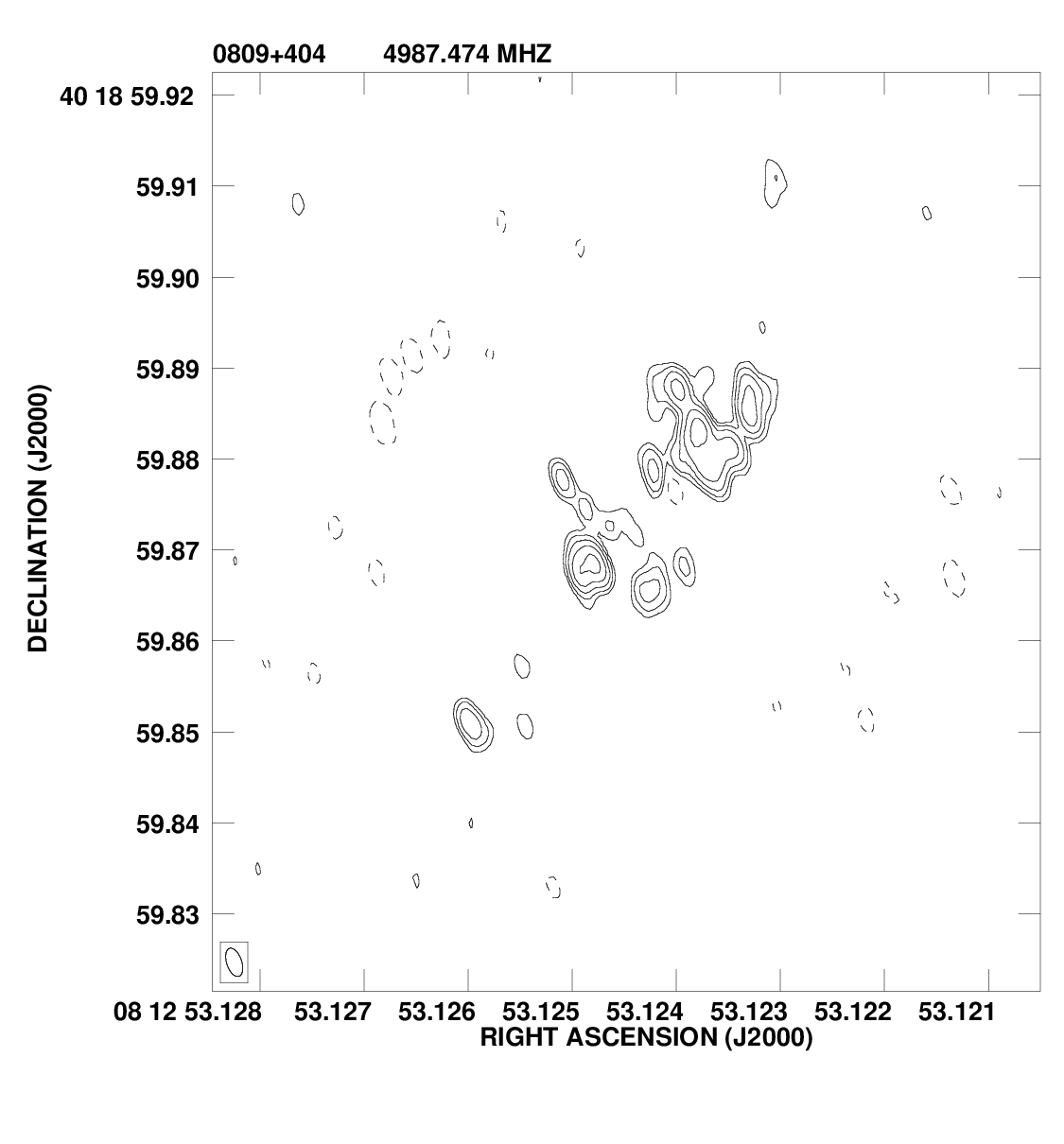}
\includegraphics[scale=0.3]{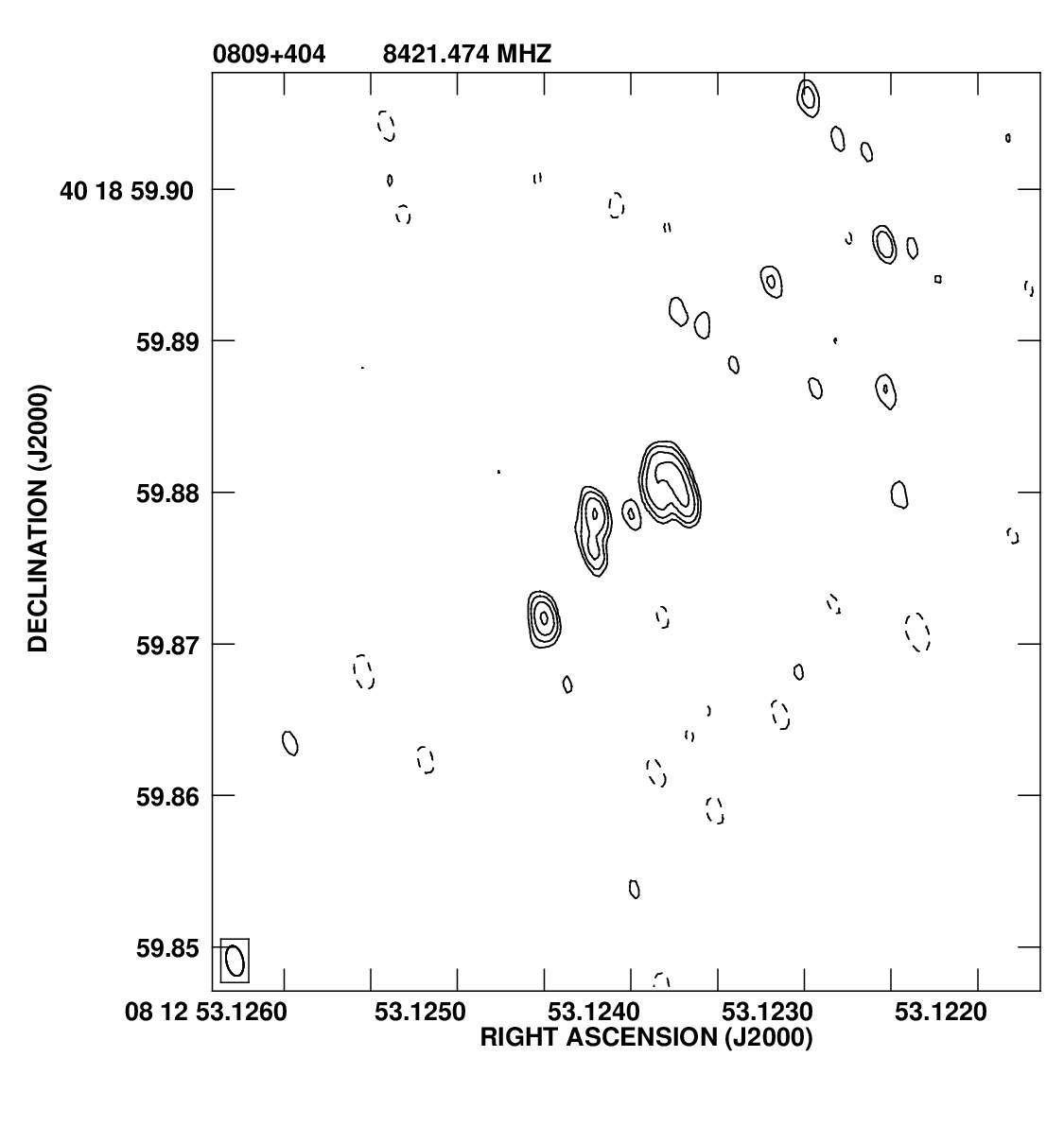}
\caption{VLBA+Effelsberg map of 0809+404 at 1.6\,GHz and VLBA 5 and 8.4\,GHz
maps. Contours increase by a factor 2 and the first contour level   
corresponds to $\approx 3\sigma$, which is 0.33\,mJy/beam for 1.6\,GHz map,
0.11\,mJy/beam for 5\,GHz map and 0.17\,mJy/beam for 8.4\,GHz map.}    
\label{0809+404_maps}
\end{figure*}

In this paper, only the results for one of the 6~sources will be
presented, namely for the galaxy 0809+404. The results of all the
observations, which seem to indicate that there does exist a class of
highly compact, but not necessarily core-dominated radio sources with
steep spectra, will be presented in a forthcoming paper. Our VLBA maps of
0809+404, show it to have a double structure (Fig.~\ref{0809+404_maps}).
However, it is important to note that VLA observations at 4.86 and
8.46\,GHz made by \citet{f00} also show a weak western component separated
from the double structure by 1\farcs3. Both components of the double
structure are fading away at the higher frequencies, so neither of
them is a core. Eventually, at 15.4\,GHz we have been unable to detect
0809+404 at all. Thus, the possibility this source might have a core-jet
structure has been excluded.
The spectral indices calculated between 1.6 and 4.9\,GHz are
very steep (Table~\ref{table2}), although it was not easy to assess them
precisely because of a break-up of the structure. There is no indication
of any hotspots. Therefore, in our opinion, 0809+404 is the {\bf best
example} of an {\bf ultra-compact fader} so far.

\section{Conclusions}

VLBI maps for 0809+404, one of 6~highly compact yet steep spectrum objects,
provide a compelling evidence that the activity of a RLAGN can cease
almost at any stage of its evolution and the length of the active phase
can span a few orders of magnitude. Such a possibility has been examined
by \citet{hat01} and \citet{jan04}. According to them, galaxies spend the
greater part of their lifetime, say $\sim$70\%, in a `quiescent' state and
$\sim$30\% in an active state and with length of the active phase of an AGN
as well as the timescale of the re-occurrence of activity being determined by
the mass of the Supermassive Black Hole (SMBH).  Specifically, if the SMBH
mass is assumed to be of the order of $10^7$ M$_\odot$ --- and such an
assumption is plausible for RLAGNs \citep{wo02,osh02} --- the length of
the activity period may be as low as $\sim10^3$ years. This means the
transition to the fader phase can also happen at a very early stage of
evolution i.e. at the CSO stage. Our discovery of ultra-compact steep
spectrum radio sources seems to be in an accordance with that model.  
Note also that such an interpretation is in agreement with an early
hypothesis on the nature of CSOs given by \citet{r94}. According to them,
some, or even the majority of CSOs, may be short-lived phenomena because
of a lack of stable, long-lasting fuelling.


\begin{table*}[t]
\caption[]{Flux densities of 0809+404 principal components at
observed frequencies}
\begin{center}
\begin{tabular}{@{}c c c r c c c@{}}
\hline
\hline
   & & & & & & \\
~~~RA & DEC & ${\rm S_{1.6GHz}}$ & \multicolumn{1}{c}{${\rm S_{4.9GHz}}$} &
$\alpha_{1.6\mathrm{GHz}}^{4.9\mathrm{GHz}}$&${\rm S_{8.4GHz}}$&
$\alpha^{8.4\mathrm{GHz}}_{4.9\mathrm{GHz}}$~~~~~~\\
~~~ h~m~s & $\degr$~$\arcmin$~$\arcsec$ & mJy &\multicolumn{1}{c}{mJy}&
&mJy&~~~~~~\\
~~~(1)& (2) & (3) & \multicolumn{1}{c}{(4)} & (5) & (6) & (7)~~~~~~\\
\hline
\hline
~~~08 12 53.123 &40 18 59.880 &155.5&12.9 &$-$2.25&6.7 &$-$1.27~~~~~~\\
~~~08 12 53.124 &40 18 59.871 &154.3&6.2  &$-$2.92&5.3 &$-$0.27~~~~~~\\
~~~08 12 53.126 &40 18 59.851 &---   &1.1
&---&---&---~~~~~~\\
\hline
\end{tabular}
\end{center}  
\label{table2}
\end{table*}


\end{document}